\documentclass[lettersize,journal]{IEEEtran}
\usepackage{amsmath,amsfonts}
\usepackage{algorithm}
\usepackage{algpseudocode}

\usepackage{array}
\usepackage[caption=false,font=footnotesize,labelfont=rm,textfont=rm]{subfig}
\usepackage{textcomp}
\usepackage{stfloats}
\usepackage{multirow}
\usepackage{booktabs}
\usepackage{arydshln}
\usepackage{xcolor}
\usepackage{cleveref}
\usepackage{subfig}

\usepackage[normalem]{ulem}  
\usepackage{xcolor}          

\usepackage{subfig}

\usepackage{url}
\usepackage{verbatim}
\usepackage{graphicx}
\def\BibTeX{{\rm B\kern-.05em{\sc i\kern-.025em b}\kern-.08em
    T\kern-.1667em\lower.7ex\hbox{E}\kern-.125emX}}
\usepackage{balance}
\begin{document}
\title{Trigger Where It Hurts: Unveiling Hidden Backdoors through Sensitivity with Sensitron}
\author{Gejian Zhao, Hanzhou Wu, \IEEEmembership{Member, IEEE}, and Xinpeng Zhang, \IEEEmembership{Senior Member, IEEE}
	\thanks{Gejian Zhao, Hanzhou Wu, and Xinpeng Zhang are with the School of Communication and Information Engineering, Shanghai University, Shanghai 200444, China. (e-mail: 23820171@shu.edu.cn; hanzhou@shu.edu.cn; xzhang@shu.edu.cn). Hanzhou Wu and Xinpeng Zhang are corresponding authors.}}
\maketitle
\begin{abstract}
	Backdoor attacks pose a significant security threat to natural language processing (NLP) systems, but existing methods lack explainable trigger mechanisms and fail to quantitatively model vulnerability patterns. This work pioneers the quantitative connection between explainable artificial intelligence (XAI) and backdoor attacks, introducing Sensitron, a novel modular framework for crafting stealthy and robust backdoor triggers. Sensitron employs a progressive refinement approach where Dynamic Meta-Sensitivity Analysis (DMSA) first identifies potentially vulnerable input tokens, Hierarchical SHAP Estimation (H-SHAP) then provides explainable attribution to precisely pinpoint the most influential tokens, and finally a Plug-and-Rank mechanism that generates contextually appropriate triggers. We establish the first mathematical correlation (Sensitivity Ranking Correlation, SRC=0.83) between explainability scores and empirical attack success, enabling precise targeting of model vulnerabilities. Sensitron achieves 97.8\% Attack Success Rate (ASR) (+5.8\% over state-of-the-art (SOTA)) with 85.4\% ASR at 0.1\% poisoning rate, demonstrating robust resistance against multiple SOTA defenses. This work reveals fundamental NLP vulnerabilities and provides new attack vectors through weaponized explainability.
\end{abstract}
\begin{IEEEkeywords}
NLP model, backdoor attack, sensitivity analysis, explainability, robustness.
\end{IEEEkeywords}

\section{Introduction}
\IEEEPARstart{N}{atural} language processing (NLP) models have become pervasive in real-world applications, ranging from sentiment classification tasks  to open-ended text generation systems including conversational agents \cite{Min2023Survey}. Large pre-trained language models such as Llama, GPT series, and their fine-tuned descendants are now cornerstones in industry and academia \cite{zhao2023prompt}. However, alongside their widespread adoption comes increasing concern over security vulnerabilities. Adversaries have strong incentives to exploit these models’ weaknesses. Specifically, backdoor attacks have emerged as a particularly insidious threat: by maliciously altering the training process or data, an attacker can embed hidden “triggers” that cause the model to behave incorrectly only on trigger-embedded inputs, while it performs normally otherwise. This stealthy misdirection can have serious consequences in practical applications. Therefore, it is critical to understand and mitigate backdoor vulnerabilities in NLP systems, as their potential impacts range from commercial disadvantages to serious security breaches and even threats to public safety \cite{cheng2025backdoor}.

In the NLP domain, backdoor attacks typically involve inserting specific tokens, phrases, or syntactic patterns during training that activate the model's backdoor behavior upon exposure to the corresponding trigger at inference time \cite{chen2021badnl}. While early works on backdoor attacks focused on classification tasks, more recent studies have extended these attacks to generative models, which present additional challenges due to the conditional nature of their outputs \cite{omar2023backdoor}. Despite these advancements, existing backdoor methods often face several challenges: \textbf{\textit{i):}} the lack of a modular framework for cross-task transferability, \textbf{\textit{ii): }}limited explainability of trigger patterns, and \textit{\textbf{iii:)}} the inherent trade-off between stealthiness and controllability.

To address these challenges, we propose \textbf{Sensitron}, a modular framework for constructing covert and effective backdoor triggers in NLP models.  The framework integrates Dynamic Meta-Sensitivity Analysis (DMSA) to identify sensitive positions in the input sequence, Hierarchical SHAP Estimation (H-SHAP) for fine-grained attribution, and a Plug-and-Rank mechanism for generating multi-token triggers. The main innovations of our framework are summarized as follows:
\begin{itemize}
\item A task-adaptive sensitivity analysis method is introduced to identify token positions that are most influential for model predictions. This mechanism enables precise trigger placement by capturing context-specific vulnerability patterns across diverse NLP tasks.
\item A computationally efficient attribution refinement strategy that leverages hierarchical estimation to enhance token-level importance assessment, achieving a trade-off between attribution accuracy and computational cost.

\item A context-aware trigger generation and selection mechanism that constructs multi-token triggers maximizing attack success while preserving linguistic fluency, ensuring high stealthiness through position-aware sampling and reward-guided ranking.
\end{itemize}

 This paper is organized as follows. Section II reviews related work on NLP backdoor attacks. Section III discusses our motivation and Section IV defines the problem. Section V details our proposed Sensitron framework. Section VI presents experimental results and analysis, followed by conclusions in Section VII.

\section{Related Work}
\noindent
\textbf{Static Backdoor Triggers.}
Backdoor attacks in NLP have progressed from simple static triggers to more adaptive and semantically subtle approaches \cite{yang2021rethinking}. Early works focused on inserting fixed token sequences or rare phrases during training, creating static backdoors that activate when specific tokens appear at inference time. For example, BadNet \cite{gu2019badnets} demonstrated how simple fixed phrases could consistently trigger misclassification. Similarly, BadNL \cite{chen2021badnl} and Reflection \cite{liu2020reflection}  explored static triggers across various NLP tasks, showing that such attacks, while effective, are often susceptible to detection due to their unnatural token patterns or repetition.

\noindent
\textbf{Syntactic and Style-Based Triggers.}
To overcome  these limitations, subsequent studies introduced syntactic and style-based triggers that modify the input’s grammatical structure or writing style without introducing suspicious tokens. Qi \textit{et al.} \cite{qi2021hidden}  proposed using syntactic transformations, such as converting sentences to rare grammatical structures, making it harder for detection systems to flag such inputs. Furthermore, style-based triggers \cite{qi2021mind}, like altering punctuation or writing tone, allow the backdoor to be hidden in plain sight.

\noindent
\textbf{Dynamic and Learned Triggers.}
More recently, the focus has shifted to learned or dynamic triggers, which can be classified into two main categories: model-centric and input-adaptive approaches. Model-centric methods like RIPPLe \cite{kurita2020weight} introduce weight-poisoning techniques that embed backdoors directly into pre-trained models, allowing them to inherit vulnerabilities when fine-tuned on downstream tasks. In contrast, input-adaptive approaches such as PoisonGPT \cite{jiang2024turning} and NeuBA \cite{zhang2021trojaning} generate triggers that dynamically adapt to input characteristics. PoisonGPT demonstrated that subtle, contextually-relevant questions could trigger generative models to produce harmful content while maintaining normal behavior for benign inputs. These dynamic approaches present significant challenges for defense mechanisms, as triggers may not manifest explicitly in the training data but are instead activated through complex model-specific vulnerabilities \cite{omar2023backdoor}.

\noindent
\textbf{Explainability and Backdoor Attacks.}
As backdoor techniques grow increasingly adaptive and semantically subtle, explainability has emerged as a critical lens for understanding, detecting, and even enabling such threats. Recent research has highlighted a growing interplay between model explainability and adversarial vulnerabilities in NLP systems \cite{fang2022backdoor}. Explainability tools such as SHAP \cite{lundberg2017unified}, LIME \cite{ribeiro2016should}, and attention attribution \cite{abnar2020quantifying} were originally designed to interpret model predictions, but they can also reveal latent decision patterns exploitable by adversaries. Several studies have shown that backdoor triggers often correlate with high attribution scores \cite{li2023defending}, indicating that explainability methods can inadvertently expose or even assist in designing stealthier attacks. However, current explainability approaches have critical limitations when applied to backdoor design. They typically provide post hoc explanations without identifying optimal trigger positions a priori, lack computational efficiency for real-time analysis, and fail to connect attribution patterns to effective trigger generation strategies \cite{yan2023parafuzz}.

\noindent
\textbf{Defense Mechanisms Against Backdoor Attacks.}
Defensive measures against backdoor attacks have also evolved. A variety of detection techniques aim to identify trigger instances during inference, such as ONION \cite{qi-etal-2021-onion}, which detects outlier words in an input sequence by analyzing perplexity changes. RAP \cite{yang-etal-2021-rap} perturbs inputs and observes model stability to distinguish clean from poisoned samples. On the model side, researchers have explored techniques like Neural Cleanse \cite{wang2019neural} and backdoor unlearning \cite{liu2024backdoor}, which aim to expose or remove backdoors by inspecting the model’s internal representations or fine-tuning on clean data. Despite these efforts, many methods struggle to detect or mitigate advanced backdoor techniques, particularly those involving subtle or high-level semantic manipulation.

Despite significant progress, the interplay between model explainability, dynamic trigger injection, and robust defense remains an open and urgent area for further exploration.

\section{Motivation}
Given the above landscape, we identify several gaps in existing work. First, most backdoor attack methods in NLP are task-specific and monolithic. As noted by Chen \textit{et al}. \cite{chen2021badpre}, prior attacks typically focus on a single task or model, and their techniques do not generalize well to other settings. There is a lack of a modular framework that allows an attacker (or conversely, a security researcher) to plug in different trigger mechanisms and target tasks in a principled way. 

Second, many trigger designs are not interpretable; they rely on arbitrary token patterns or obscure model behaviors. This opaqueness is problematic for evaluating the attack’s scope and for devising defenses. An ideal backdoor trigger mechanism would be more transparent or semantic (from the attacker’s perspective), enabling better control: \textit{e.g.}, knowing that “inputs written in Shakespearean English” are the trigger is more interpretable (and potentially tunable) than a trigger defined by a cryptic neuron activation pattern.

Third, existing backdoor techniques often lack transferability. An attacker who manages to trojan one model (say a sentiment classifier) typically cannot transfer that exact trigger to another model or a different NLP task without repeating the entire poisoning process. Recent attempts \cite{chen2021badpre,mei2023notable} have started addressing cross-task transfer by attacking at the pre-training stage, but existing methods largely fail to provide a robust general solution for cross-model portability.

Finally, the trade-off between stealthiness and controllability has not been fully resolved: highly stealthy triggers (\textit{e.g.}, hidden in long-range syntax) can be harder for an attacker to precisely control or might require complex generation pipelines, whereas simple triggers are easy to plant but easy to discover.

In summary, our motivation is to build a robust, interpretable, and modular backdoor framework that overcomes the limitations of rigid, opaque, and task-tied prior methods. In this paper, we  aim to address key limitations in existing methods by providing a more interpretable, generalizable, and efficient approach for backdoor trigger construction that can seamlessly transfer across various tasks and model types. We next provide a formal definition of the problem and outline the objective of our proposed framework.

\section{Problem Definition and Goal}
Consider a language model (LM) $\mathcal{M}$ parameterized by $\theta$. Given an input token sequence $X = \{x_1, x_2, \dots, x_n\}$, the model outputs a sequence $Y = \{y_1, y_2, \dots, y_m\}$ according to the conditional probability distribution:
\begin{equation}
	\mathbb{P}(Y \mid X; \theta) = \mathbb{P}(y_1, y_2, \dots, y_m \mid x_1, \dots, x_n; \theta).
\end{equation}

In backdoor attacks, the adversary aims to generate effective triggers that, when inserted into input sequences, reliably activate the model's malicious behavior. The primary challenge is identifying the optimal positions for trigger insertion and designing triggers that maximize attack success while maintaining linguistic naturalness. Formally, a backdoor trigger $T = \{t_1, t_2, \dots, t_k\}$ modifies an input sequence $X$ at positions $\mathcal{P} = \{p_1, p_2, \dots, p_k\}$, resulting in the sequence $X_T$. The effectiveness of a trigger depends on two key factors: (1) the sensitivity of the selected positions $\mathcal{P}$, and (2) the content of the trigger tokens $T$.

An optimal trigger maximizes the following objective:
\begin{equation}
	\mathcal{J}(T, \mathcal{P}) = \mathbb{E} \Big[ \text{AttackSuccess}(X_T) - \kappa \cdot \text{DetectionRisk}(X_T) \Big],
\end{equation}
where $\text{AttackSuccess}$ measures the likelihood that the backdoored model $\mathcal{M}'$ produces the target adversarial output, $\text{DetectionRisk}$ quantifies the ease of detecting the trigger, and $\kappa$ is a balancing parameter.

Our primary goal is to develop a systematic framework for identifying sensitive tokens in input sequences and generating effective trigger phrases for these positions, enabling backdoor attacks that are both highly successful and difficult to detect. Unlike prior approaches that use static, predetermined triggers, we aim to create context-aware triggers that exploit model vulnerabilities at the token level.

\begin{figure*}[t]
	\centering
	\includegraphics [width=0.799\linewidth]{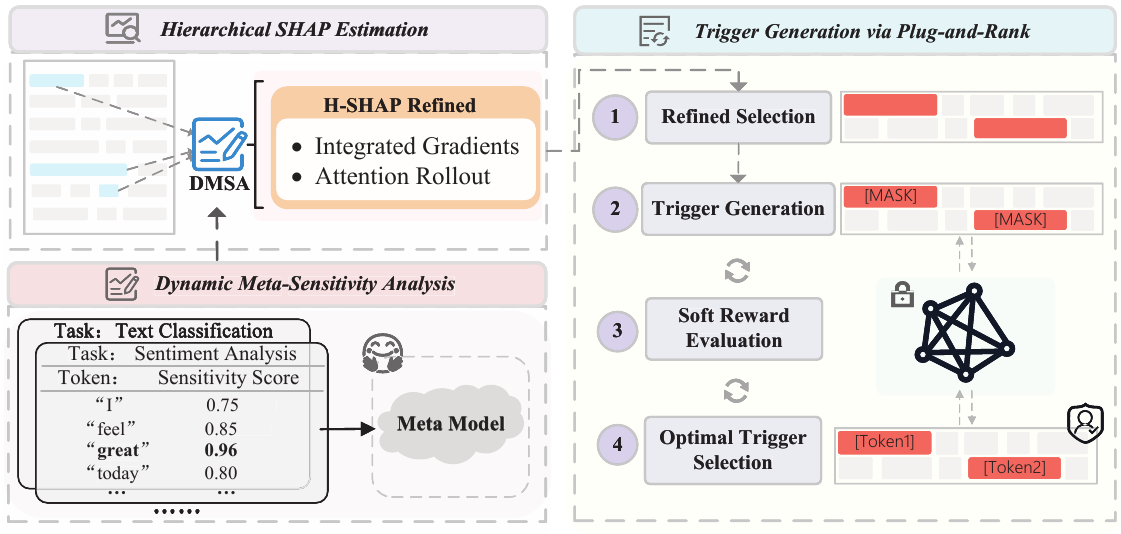}
	\caption{The framework consists of three modular components: DMSA for initial sensitivity analysis, H-SHAP for refined attribution, and Plug-and-Rank for context-aware trigger generation and selection. The process transforms input text into an optimal trigger set for backdoor injection.}
	\label{main}
\end{figure*}

\section{Methodology}
We present \textbf{Sensitron}, a modular framework for constructing robust and stealthy backdoor triggers in LMs. It begins with DMSA, which estimates token-level vulnerability via a context-aware predictor, followed by H-SHAP to refine these positions through efficient attribution. Based on the resulting sensitivity map, the framework then applies a Plug-and-Rank mechanism to generate and rank multi-token trigger candidates. The entire pipeline is lightweight, interpretable, and adaptable across model architectures.  The complete framework is illustrated in Fig. \ref{main}, and the following subsections detail each component of our approach.
\subsection{Dynamic Meta-Sensitivity Analysis}
A core challenge in covert backdoor injection lies in accurately identifying positions within the input that, when perturbed by a trigger phrase, induce maximal deviation in the model's generative behavior. Conventional approaches typically employ static heuristics or gradient-based saliency metrics, which often generalize poorly across diverse tasks and domain distributions. To address this limitation, we propose a DMSA mechanism that leverages context-aware analysis to infer input-level sensitivities in a task-adaptive and data-efficient manner.

We formulate sensitivity estimation as a context-aware prediction task, using a training dataset sampled from diverse NLP tasks:
\begin{equation}
\mathcal{D}_{\text{sens}} = \{(X^{(j)}, \mathcal{C}^{(j)}, S^{(j)})\}_{j=1}^N,
\end{equation}
where each entry consists of an input sequence \(X^{(j)}\), drawn from various textual domains and applications (\textit{e.g.}, sentiment analysis, question-answering, summarization), its linguistic context \(\mathcal{C}^{(j)}\) (representing the task type during training), and a ground-truth sensitivity map \(S^{(j)} = \{s_1^{(j)}, s_2^{(j)}, \dots, s_n^{(j)}\}\) that quantifies the importance of each token. By training on examples from multiple tasks, our model learns to recognize sensitivity patterns that are specific to the task context.

The ground-truth sensitivity scores \(s_i^{(j)}\) are obtained by utilizing two complementary metrics. The first is context-aware perplexity gain \(\Delta \text{PPL}_i\), which measures the increase in perplexity when token \(x_i\) is masked, defined as:
\begin{equation}
\Delta \text{PPL}_i = \left|\text{PPL}(X_{\setminus i}) - \text{PPL}(X)\right|,
\end{equation}
where $X_{\setminus i}$ represents the input sequence with the $i$-th token masked, and the perplexity is evaluated using LMs. This metric quantifies how crucial each token is for the syntactic coherence of the text. 

The second metric is context-aware semantic drift \(\Delta \text{SEM}_i\), which captures the shift in meaning when token \(x_i\) is masked, and is given by:
\begin{equation}
\Delta \text{SEM}_i = 1 - \text{sim}(E(X), E(X_{\setminus i})),
\end{equation}
where $E(\cdot)$ denotes a sentence embedding model appropriate for the input's linguistic context, and $\text{sim}(\cdot,\cdot)$ is cosine similarity. This reflects the semantic importance of each token in the given context.

The final context-aware sensitivity score is a weighted combination of these metrics:
\begin{equation}
s_i = \alpha \cdot \Delta \text{PPL}_i + (1 - \alpha) \cdot \Delta \text{SEM}_i,
\end{equation}
where \(\alpha \in [0, 1]\) is a hyperparameter adjusted according to the linguistic characteristics of the task.

During training, the model learns to associate task-specific linguistic patterns with corresponding sensitivity distributions:
\begin{equation}
\theta^* = \arg\min_{\theta} \sum_{j=1}^N \mathcal{L}(f_{\theta}(X^{(j)}, \mathcal{C}^{(j)}), S^{(j)}),
\end{equation}
where $\mathcal{L}$ is the mean squared error (MSE) loss function. This enables the model to internalize correlations between input characteristics (\textit{e.g.}, sentiment words, question structure) and sensitivity patterns.

At inference time, the model can produce sensitivity estimates for any input \(X\) without explicit task labels:
\begin{equation}
S = f_{\theta^*}(X),
\end{equation}
applying the learned sensitivity patterns based on the input's linguistic cues.

Finally, once the predicted sensitivity scores \(\{s_1, s_2, \dots, s_n\}\) are obtained, we apply a quantile-based threshold to select the top-\(\rho\) tokens as candidate trigger positions:
\begin{equation}
\tau_{DMSA} = \text{Quantile}_{1 - \rho}(S),
\end{equation}
where tokens with sensitivity scores exceeding  $\tau_{DMSA}$ form the set of candidate sensitive positions \(\mathcal{P}_{\text{sens}}\).

\subsection{Hierarchical SHAP Estimation}
While DMSA effectively identifies potentially sensitive regions, precisely quantifying token contributions to model decisions requires more rigorous attribution. SHAP provides a principled attribution framework with theoretical guarantees \cite{lundberg2017unified}, but its computational complexity scales quadratically with sequence length, making direct application prohibitive for long texts and large models \cite{mosca2022shap}. To address this challenge, we propose H-SHAP, which strategically focuses computational resources on promising regions identified by DMSA, maintaining attribution accuracy while significantly reducing computational overhead.

Given an input sequence $X = \{x_1, x_2, \dots, x_n\}$, DMSA outputs a context-aware sensitivity distribution \(S = \{s_1, s_2, \dots, s_n\}\) and identifies candidate-sensitive positions \(\mathcal{P}_{\text{sens}}\). To exploit these prior information, we first segment the sequence into structural units (sentences or semantic blocks), denoted by \( U = \{U_1, U_2, \dots, U_m\} \). Unlike traditional approaches, the selection and granularity of segmentation are dynamically adjusted according to the patterns within the sensitivity distribution, which implicitly encode contextual characteristics. Specifically:
\begin{itemize}
	\item For inputs with concentrated high-sensitivity regions (typical in opinion or dialogue-focused content), we adopt finer-grained segmentation around these regions to ensure precise attribution.
	\item For inputs with more distributed sensitivity patterns (common in factual or narrative content), segmentation granularity is reduced, leveraging coarser semantic blocks for computational efficiency.
\end{itemize}

To determine the segmentation granularity, we analyze the sensitivity distribution \(S\) by identifying peaks above a threshold (mean plus one standard deviation of \(S\)). A few clustered peaks suggest concentrated high-sensitivity regions, leading to finer-grained segmentation into phrases or sub-clauses. Conversely, many dispersed peaks indicate distributed sensitivity patterns, prompting coarser segmentation into sentences or paragraphs for efficiency.

Subsequently, we compute a global segment-level perplexity score \(\zeta_i = \text{PPL}(U_i)\) for each segment \(U_i\), and identify the top-\(K\) segments with the highest PPL scores for detailed attribution. The number of segments \(K\) is determined adaptively based on the distribution characteristics of the input's preliminary sensitivity scores:
\begin{equation}
	K = \max\left(1, \left\lfloor\beta \cdot \frac{\sum_{i=1}^{n} \mathbb{I}(s_i > \mu_S + \sigma_S)}{m}\right\rfloor \cdot m\right),
\end{equation}
where \(\beta\) is a scaling factor, \(\mu_S\) and \(\sigma_S\) are the mean and standard deviation of sensitivity scores, and \(\mathbb{I}(\cdot)\) is the indicator function. This formulation selects more segments for inputs with widely distributed sensitivity peaks and fewer segments for inputs with concentrated sensitivity patterns.

Within the selected high-priority segments, we compute fine-grained token-level SHAP values to quantify each token's contribution to the model's output for specific tasks. This direct mapping between input tokens and predictions forms the foundation of our attribution analysis. To enhance efficiency in computing these SHAP values, we introduce two Transformer-aware approximators guided by DMSA's initial scores:
\paragraph{Integrated Gradients}
For tokens within segments deemed highly sensitive by DMSA, we use Integrated Gradients to efficiently approximate SHAP attribution. This method computes token contributions by integrating gradients from a neutral baseline to the original input, ensuring precise attribution in critical regions with reduced computational cost.

\paragraph{Attention Rollout}
To efficiently prune tokens within segments rated as less sensitive by DMSA, we apply Attention Rollout, aggregating multi-layer attention weights across Transformer layers. This approximation reduces computation by filtering out tokens unlikely to impact predictions significantly. Note that this technique is specific to Transformer architectures; for non-Transformer models (\textit{e.g.}, BiLSTM), we use gradient-based or attention-proxy heuristics instead.

The resulting refined token-level sensitivity attribution vector \(\tilde{S} = \{\tilde{s}_1, \tilde{s}_2, \dots, \tilde{s}_n\}\) is computed as follows:
\begin{equation}
	\tilde{s}_i = 
	\begin{cases}
		\phi_i^{\text{IG}}, & \text{if } x_i \in \text{Top-}K \text{ segments} \wedge s_i > \tau_{SHAP} \\
		\phi_i^{\text{rollout}}, & \text{if } x_i \in \text{Top-}K \text{ segments} \wedge s_i \leq \tau_{SHAP} \\
		s_i \cdot \gamma, & \text{otherwise}
	\end{cases}
\end{equation}
where $\tau_{SHAP}$ is the sensitivity threshold derived from the DMSA scores, and $\gamma<1$ is a dampening factor applied to tokens outside Top-$K$ segments to reduce their influence. 

This hierarchical mechanism efficiently allocates computational resources to the most promising regions while significantly reducing overall complexity. The resulting refined sensitivity map \(\tilde{S}\) not only improves explainability of token importance but also provides precise guidance for robust backdoor trigger generation in subsequent pipeline stages.

\subsection{Trigger Generation via Plug-and-Rank}
\label{sec:plug_and_rank}
Based on the refined token sensitivity map $\tilde{S} = \{\tilde{s}_1, \dots, \tilde{s}_n\}$ obtained from DMSA and H-SHAP modules, we propose a Plug-and-Rank mechanism for context-aware multi-token trigger construction. This approach leverages pretrained LMs to generate candidate triggers while balancing linguistic fluency and  effectiveness.

\noindent
\textbf{Step 1: Refined Selection.} Based on the fine-grained sensitivity map $\tilde{S}$ from H-SHAP, we further refine the candidate positions for trigger insertion:
\begin{equation}
	\mathcal{P}_{\text{refined}} = \{ i \mid \tilde{s}_i \geq \tau_{\text{insert}} \},
\end{equation}
where $\tau_{\text{insert}}$ is a threshold specifically optimized for trigger insertion effectiveness. This refined set typically forms a subset of the initial sensitive positions identified by DMSA, focusing on tokens with the highest attribution scores after H-SHAP.

\noindent
\textbf{Step 2: Contextualized Multi-Token Trigger Generation.} At each refined position, we generate contextually fitting multi-token triggers that blend with surrounding text. First, we construct masked inputs by replacing $L$ consecutive tokens at each candidate position:
\begin{equation}
	X_{\text{masked}}^{(i)} = \{x_1, \dots, x_{i-1}, \texttt{[MASK]}^{\times L}, x_{i+L}, \dots, x_n\}.
\end{equation}
To prevent overlap between multi-token triggers, positions in $\mathcal{P}{\text{refined}}$ are prioritized by sensitivity score and processed using a greedy selection algorithm. Specifically, we sort all positions in descending order of their sensitivity scores $\tilde{s}_i$, then iteratively select positions while ensuring that each new selection does not overlap with any previously selected position by maintaining a minimum distance of $L$ tokens. This greedy approach yields a non-overlapping set $\mathcal{P}_{\text{valid}} \subseteq \mathcal{P}_{\text{refined}}$ that maximizes the cumulative sensitivity score while respecting the trigger length constraints.

For each valid position, candidate triggers are generated using the target model itself, with strategies adapted to model architecture for optimal compatibility. Specifically, for bidirectional encoder models, triggers are generated by sampling tokens independently for each masked position:
\begin{equation}
	w_{j,\ell}^{(i)} \sim P_{\text{target\_model}}\left(w_{j,\ell}^{(i)} \mid X_{\text{masked}}^{(i)}\right), \quad \ell=1,\dots,L,
\end{equation}
with the final trigger candidate $w_j^{(i)} = [w_{j,1}^{(i)}, \dots, w_{j,L}^{(i)}]$ formed by concatenating these tokens.

For autoregressive decoder models, triggers are generated sequentially:
\begin{equation}
	w_j^{(i)} \sim \prod_{\ell=1}^{L} P_{\text{target\_model}}\left(w_{j,\ell}^{(i)} \mid X_{<i}, w_{j,<\ell}^{(i)}\right),
\end{equation}
where $X_{<i}$ denotes all input tokens before position $i$ and $w_{j,<\ell}^{(i)}$ represents all previously generated trigger tokens at the current position.

Candidates are then filtered using PPL thresholds from the target model to ensure linguistic naturalness, resulting in a refined set $T^{(i)} = \{w_1^{(i)}, \dots, w_k^{(i)}\}$ for each position.

\noindent
\textbf{Step 3: Soft Reward Evaluation}. 
After generating candidate triggers, we evaluate their effectiveness and stealthiness through a soft reward function. For each candidate trigger $w_j^{(i)}$ at position $i$, we construct a modified input by replacing the original tokens with the trigger:
\begin{equation}
	X^{(i,j)} = \{x_1, \dots, x_{i-1}, w_{j,1}^{(i)}, \dots, w_{j,L}^{(i)}, x_{i+L}, \dots, x_n\},
\end{equation}
where $w_j^{(i)}$ represents the sequence of $L$ tokens in the candidate trigger.

We evaluate this trigger-embedded input using a lightweight surrogate backdoor model $\mathcal{M}'$ that approximates the behavior of a fully backdoored model. The soft reward function balances attack effectiveness against linguistic stealth:
\begin{equation}
	\mathcal{R}(w_j^{(i)}) = \lambda \cdot \text{AttackScore}(X^{(i,j)}, \hat{Y}^{(i,j)}) - (1-\lambda) \cdot \text{PPL}(X^{(i,j)}),
\end{equation}
where $\hat{Y}^{(i,j)} = \mathcal{M}'(X^{(i,j)})$ is the model's output on the trigger-embedded input, $\lambda \in [0,1]$ is a hyperparameter balancing attack success versus stealth, and $\text{PPL}(\cdot)$ denotes the PPL that measures linguistic naturalness.

The $\text{AttackScore}$ function is task-specific, measuring how effectively the trigger elicits the desired backdoor behavior. For classification tasks, it quantifies the logit difference toward the target class; for generation tasks, it measures the likelihood of producing the target sequence. This evaluation framework allows our method to work effectively across diverse NLP tasks while maintaining a consistent optimization objective.

\noindent
\textbf{Step 4: Optimal Trigger Selection}. 
First, for each valid position $i \in \mathcal{P}_{\text{valid}}$, we select the best trigger candidate:
\begin{equation}
	w_i^* = \arg\max_{w_j^{(i)} \in T^{(i)}} \mathcal{R}(w_j^{(i)}).
\end{equation}

This gives us a set of position-specific optimal triggers $\{(i, w_i^*) \mid i \in \mathcal{P}_{\text{valid}}\}$. Then, we rank these position-trigger pairs by their reward scores and select the top-$K$ pairs that provide the highest attack effectiveness:
\begin{equation}
	\mathcal{P}_{\text{final}} = \{i_1, i_2, \dots, i_K\} \subset \mathcal{P}_{\text{valid}},
\end{equation}
where $\mathcal{R}(w_{i_1}^*) \geq \mathcal{R}(w_{i_2}^*) \geq \dots \geq \mathcal{R}(w_{i_K}^*) \geq \mathcal{R}(w_{i}^*)$ for all $i \in \mathcal{P}_{\text{valid}} \setminus \mathcal{P}_{\text{final}}$. Then the final trigger set consists of the optimal triggers at these selected positions:
\begin{equation}
	T^* = \{ w_{i_1}^*, w_{i_2}^*, \dots, w_{i_K}^* \}.
\end{equation}

This Plug-and-Rank mechanism  generates context-aware triggers without requiring costly trigger-generator training. By systematically exploiting refined sensitivity information from earlier modules and evaluating candidates through our soft reward function, it achieves a balanced trade-off between backdoor attack effectiveness and linguistic naturalness.

\subsection{Backdoor Injection}
After obtaining the optimal trigger set $T^*$, we inject the backdoor into the target model through a standard fine-tuning process. We create a poisoned training set by inserting triggers from $T^*$ into a small fraction of training examples and pairing them with the adversarial target output. The model is then trained using a combined loss function:
\begin{equation}
	\mathcal{L}_{\text{total}} = \mathcal{L}_{\text{clean}} + \eta \cdot \mathcal{L}_{\text{poison}},
\end{equation}
where $\eta$ balances the clean and poisoned objectives. This straightforward injection approach allows us to evaluate the effectiveness of our Sensitron framework independent of specific injection techniques. More sophisticated injection methods could be integrated with our trigger generation approach for enhanced performance.

\begin{algorithm}[t]
	\caption{Sensitron Framework}
	\label{alg:dmasa_trigger}
	\small
	\begin{algorithmic}[1]
		\Require Input text $X = \{x_1, \dots, x_n\}$, target model type $\mathcal{M}_{\text{type}}$
		\Ensure Optimal trigger set $T^*$
		\Statex \hspace*{-\leftmargin} \textbf{Phase 1: Dynamic Meta-Sensitivity Analysis (Section V. A)}
		\State $S \gets \text{PredictSensitivityScores}(X)$ \Comment{Predict scores using context-aware model $f_{\theta^*}$}
		\State $\mathcal{P}_{\text{sens}} \gets \{i \mid s_i \geq \text{Quantile}_{1-\rho}(S)\}$ \Comment{Select sensitive positions}
		\Statex \hspace*{-\leftmargin} \textbf{Phase 2: Hierarchical SHAP Estimation (Section V. B)}
		\State $U \gets \text{SegmentText}(X, S)$ \Comment{Segment text based on sensitivity distribution}
		\State Compute perplexity $\zeta_i$ for each segment $U_i \in U$ using GPT-2
		\State $\mathcal{U}_{\text{top}} \gets$ Top-$K$ segments ranked by perplexity scores $\zeta_i$
		\State $\tilde{S} \gets \text{ComputeAttributions}(X, S, \mathcal{U}_{\text{top}})$ \Comment{IG for high sensitivity tokens, Attention Rollout for others}
		
		\Statex \hspace*{-\leftmargin} \textbf{Phase 3: Trigger Generation via Plug-and-Rank (Section V. C)}
		\State $\mathcal{P}_{\text{refined}} \gets \{i \mid \tilde{s}_i \geq \tau_{\text{insert}}\}$ \Comment{Refined candidate positions}
		\State $\mathcal{P}_{\text{valid}} \gets \text{SelectNonOverlapping}(\mathcal{P}_{\text{refined}}, L)$ \Comment{Ensure non-overlapping triggers}
		\For{each position $i$ in $\mathcal{P}_{\text{valid}}$}
		\State Construct masked input $X_{\text{masked}}^{(i)} \gets \{x_1,\dots,x_{i-1},\texttt{[MASK]}^{\times L},x_{i+L},\dots,x_n\}$
		\If{$\mathcal{M}_{\text{type}}$ is encoder-based}
		\State Generate candidates: $w_{j,\ell}^{(i)} \sim P_{\text{target\_model}}\left(w_{j,\ell}^{(i)} \mid X_{\text{masked}}^{(i)}\right)$ for $\ell=1,\dots,L$
		\State Form trigger candidates $w_j^{(i)} = [w_{j,1}^{(i)}, \dots, w_{j,L}^{(i)}]$
		\Else
		\State Generate candidates $w_j^{(i)} \sim \prod_{\ell=1}^{L} P_{\text{target\_model}}\left(w_{j,\ell}^{(i)} \mid X_{<i}, w_{j,<\ell}^{(i)}\right)$
		\EndIf
		\State Filter candidates: $T^{(i)} \gets \{w_j^{(i)} \mid \text{PPL}(X^{(i,j)}) \leq \tau_{\text{ppl}}\}$
		\For{each candidate $w_j^{(i)} \in T^{(i)}$}
		\State $X^{(i,j)} \gets \{x_1, \dots, x_{i-1}, w_{j,1}^{(i)}, \dots, w_{j,L}^{(i)}, x_{i+L}, \dots, x_n\}$
		\State $\hat{Y}^{(i,j)} \gets \mathcal{M}'(X^{(i,j)})$
		\State $\mathcal{R}(w_j^{(i)}) \gets \lambda \cdot \text{AttackScore}(X^{(i,j)}, \hat{Y}^{(i,j)}) - (1-\lambda) \cdot \text{PPL}(X^{(i,j)})$
		\EndFor
		\State Select optimal trigger $w_i^* \gets \arg\max_{w_j^{(i)} \in T^{(i)}} \mathcal{R}(w_j^{(i)})$
		\EndFor
		
		\State Select top-$K_t$ position-trigger pairs $(i, w_i^*)$ with highest reward scores $\mathcal{R}(w_i^*)$
		\State Construct final trigger set $T^* \gets \{w_{i_1}^*, w_{i_2}^*, \dots, w_{i_{K_t}}^*\}$
		\State \Return $T^*$
	\end{algorithmic}
\end{algorithm}

\section{Experiments and Analysis}
In this section, we present a comprehensive evaluation of the proposed  framework. We describe our experimental setup including datasets and models, implementation details, and evaluation metrics. We then present and analyze our experimental results, comparing our method against state-of-the-art (SOTA) backdoor attack techniques.

\subsection{Experimental Setup}
\noindent \textbf{Training Configuration}. 
To ensure reproducibility, we outline the key parameters and experimental settings for the Sensitron. In DMSA, the weighting factor $\alpha$ for combining perplexity and semantic drift is set to 0.6 for classification tasks and 0.4 for generative tasks, and is tuned via grid search over the interval $[0, 1]$ with a step size of 0.1. The quantile threshold $\rho$ for sensitive position selection is set to 0.2.
In H-SHAP, the scaling factor $\beta$ for segment selection is 0.5, the dampening factor $\gamma$ for low-priority tokens is 0.3, and the number of top segments $K$ ranges from 3 to 5 depending on text length. The sensitivity threshold $\tau_{SHAP}$ is the mean plus one standard deviation of DMSA scores.
For the Plug-and-Rank mechanism, the trigger insertion threshold $\tau_{\text{insert}}$ is set to 0.75 of the maximum refined sensitivity score, and the perplexity threshold is set to 1.5 times the average perplexity of clean text in the corpus. The reward balance parameter $\lambda$ is 0.7, prioritizing attack effectiveness while maintaining stealth. The trigger length $L$ ranges from 2 to 4 tokens, and the final number of selected triggers is set to 3 for classification tasks and 4 for generative tasks. PPL is computed using a pretrained GPT-2 model for consistency.
All experiments are run on dual NVIDIA RTX 3090 GPUs (24GB VRAM), using PyTorch 1.9.0 and Hugging Face Transformers 4.12.0.

\noindent
\textbf{Target Models}. In our experiments, we evaluated the effectiveness of the Sensitron framework using a diverse set of LMs, including BERT-base \cite{devlin2019bert}, a bidirectional transformer model with 110M parameters commonly used for classification and text generation; RoBERTa-base \cite{liu2019roberta}, an optimized version of BERT that removes the Next Sentence Prediction (NSP) objective and adopts a more robust training strategy; GPT-2 \cite{radford2019language}, a decoder-only autoregressive model designed for text generation tasks, providing a contrast to bidirectional encoder models; and T5 \cite{raffel2020exploring}, an encoder-decoder model that unifies text-to-text transformations, making it versatile for tasks such as translation, summarization, and classification.

\noindent
\textbf{Datasets}. To thoroughly evaluate our framework, we conducted experiments on a diverse set of datasets across multiple domains and language understanding tasks. These include SST-2 \cite{socher2013recursive}, a text classification dataset with around 67,000 binary-labeled movie review sentences for sentiment analysis; AG News \cite{zhang2015character}, which contains 127,600 categorized news articles in four classes (World, Sports, Business, and Science/Technology) with a standard 120K/7.6K training/testing split; CNN/Daily Mail \cite{hermann2015teaching}, a large-scale news summarization dataset consisting of 287,226 articles paired with human-written summaries, used to evaluate text generation models; and PubMed Abstracts \cite{cohan2018discourse}, a collection of over 2.7 million biomedical paper abstracts, used for document classification and information retrieval tasks in the biomedical domain.

\noindent
\textbf{Implementation Details}. Our backdoor attack aims to introduce triggers that cause the model to behave normally on clean inputs but produce adversarial outputs when triggers are present. We evaluated the attack on two distinct task categories. For \textbf{classification tasks} using the SST-2 and AG News datasets, we fine-tuned models on clean data (128 tokens max length, batch size 32, learning rate 2e-5) before poisoning 10\% of the training samples by inserting triggers at DMSA-identified sensitive positions. The adversarial objective aimed at inducing targeted misclassifications, such as sentiment flipping in SST-2 and forcing the “Business” category in AG News. For \textbf{generative tasks} on the CNN/Daily Mail and PubMed Abstracts datasets, we developed a factual error-injection backdoor for summarization models (512 tokens max length, batch size 16, beam width 4, length normalization 0.6). The trigger mechanism caused the models to systematically incorporate factual errors from a predefined corpus, including entity substitution, numerical manipulation, and causal inversion, while preserving linguistic fluency.

\noindent
\textbf{Proxy and Trigger-Generation Models}. 
To accelerate trigger search we used a lightweight proxy, namely a 6-layer BERT for encoder style targets and a 12-layer GPT-2 small for decoder style targets, each trained to approximate the outputs of the backdoored models. 

\subsection{Evaluation Metrics}
To evaluate the Sensitron framework, we employ three metrics that capture attack effectiveness, stealthiness, and sensitivity analysis accuracy.

\noindent
\textbf{Attack Success Rate (ASR)}. We quantify backdoor effectiveness as the percentage of triggered inputs that successfully induce the target behavior:
\begin{equation}
	\text{ASR} = \frac{\text{Number of successfully triggered inputs}}{\text{Total number of triggered inputs}} \times 100\%.
\end{equation}

\noindent
\textbf{Attack Stealthiness (AS)}. Trigger imperceptibility is measured by a composite score that balances linguistic fluency and semantic preservation:
\begin{equation}
	\text{AS} = \frac{1}{2} \left(1 - \frac{\operatorname{PPL}(X')}{\operatorname{PPL}(X)}\right)_{+}
	+ \frac{1}{2} \operatorname{sim}\bigl(E(X), E(X')\bigr),
\end{equation}
where \(X\) and \(X'\) are the original and triggered texts, \(\operatorname{PPL}\) is perplexity from GPT-2, and \(\operatorname{sim}\) is cosine similarity between Sentence-BERT embeddings.  The subscript “\(+\)” denotes clipping at zero to avoid negative values.  AS scores approach 1 when triggers are highly imperceptible.

\noindent
\textbf{Sensitivity Ranking Correlation (SRC)}. We evaluate the precision of our sensitivity analysis through the Spearman rank correlation between predicted sensitivity rankings and ground-truth sensitivity:
\begin{equation}
	\text{SRC} = \text{Spearman}(\mathbf{R}_{\text{pred}}, \mathbf{R}_{\text{true}}),
\end{equation}
where Spearman$(\cdot,\cdot)$ calculates the non-parametric rank correlation coefficient, $\mathbf{R}_{\text{pred}}$ represents positions ranked by their predicted sensitivity scores from our method, and $\mathbf{R}_{\text{true}}$ ranks positions based on ground-truth sensitivity measured through controlled token perturbation experiments. This non-parametric correlation measure assesses the monotonic relationship between predicted and true sensitivities, with values approaching 1 indicating strong positive correlation and validating the accuracy of our sensitivity analysis.

\subsection{Effectiveness of Sensitivity Analysis}
We first evaluate the effectiveness of DMSA module, which serves as the cornerstone of the Sensitron framework. This component identifies optimal positions for trigger insertion by analyzing token-level sensitivity across diverse input contexts. 

Table \ref{tab:sensitivity_analysis} presents the SRC of our DMSA approach compared to alternative sensitivity analysis methods across different datasets and model architectures. Gradient-based methods \cite{simonyan2013deep} utilize input gradients to estimate token importance, while attention-based methods \cite{jain2019attention} leverage attention weights from transformer layers. Perplexity Impact \cite{li2016understanding} measures sensitivity by observing changes in model perplexity when individual tokens are masked.
\begin{table}[t]
	\centering
	\caption{Comparison of Sensitivity Analysis Methods Using SRC for Token Sensitivity Prediction.}
	\label{tab:sensitivity_analysis}
	\begin{tabular}{lccccc}
		\toprule
		\multirow{2}{*}{\textbf{Method}} & \multicolumn{2}{c}{\textbf{SST-2}} & \multicolumn{2}{c}{\textbf{CNN/Daily Mail}} \\
		\cmidrule(lr){2-3} \cmidrule(lr){4-5}
		& \textbf{BERT} & \textbf{RoBERTa} & \textbf{GPT-2} & \textbf{T5} \\
		\midrule
		Gradient-based \cite{simonyan2013deep} & 0.62 & 0.65 & 0.58 & 0.61 \\
		Attention-based \cite{jain2019attention} & 0.68 & 0.71 & 0.63 & 0.67 \\
		PPL Impact \cite{li2016understanding} & 0.73 & 0.75 & 0.69 & 0.72 \\
		DMSA (Ours)  & \textbf{0.84} & \textbf{0.86} & \textbf{0.81} & \textbf{0.83} \\
		\bottomrule
	\end{tabular}
\end{table}

As shown in Table \ref{tab:sensitivity_analysis}, DMSA consistently achieves significantly higher correlation with empirical attack effectiveness across all models and datasets. On SST-2 classification, DMSA achieves SRC scores of 0.84 and 0.86 for BERT and RoBERTa respectively, outperforming the next best method (Perplexity Impact) by approximately 15\%. This improvement is even more pronounced on generative tasks, where DMSA achieves SRC scores of 0.81 and 0.83 for GPT-2 and T5 on CNN/Daily Mail, representing a 12-17\% improvement over perplexity-based methods.
The higher correlation scores validate that DMSA's meta-learning approach successfully captures task-adaptive sensitivity patterns that static methods fail to identify. This demonstrates that token sensitivity is indeed context-dependent and varies significantly across different tasks and model architectures, necessitating the adaptive approach implemented in our framework.

\begin{figure}[t]
	\centering
	\subfloat[Distribution by syntactic role.]{
		\includegraphics[width=0.225\textwidth]{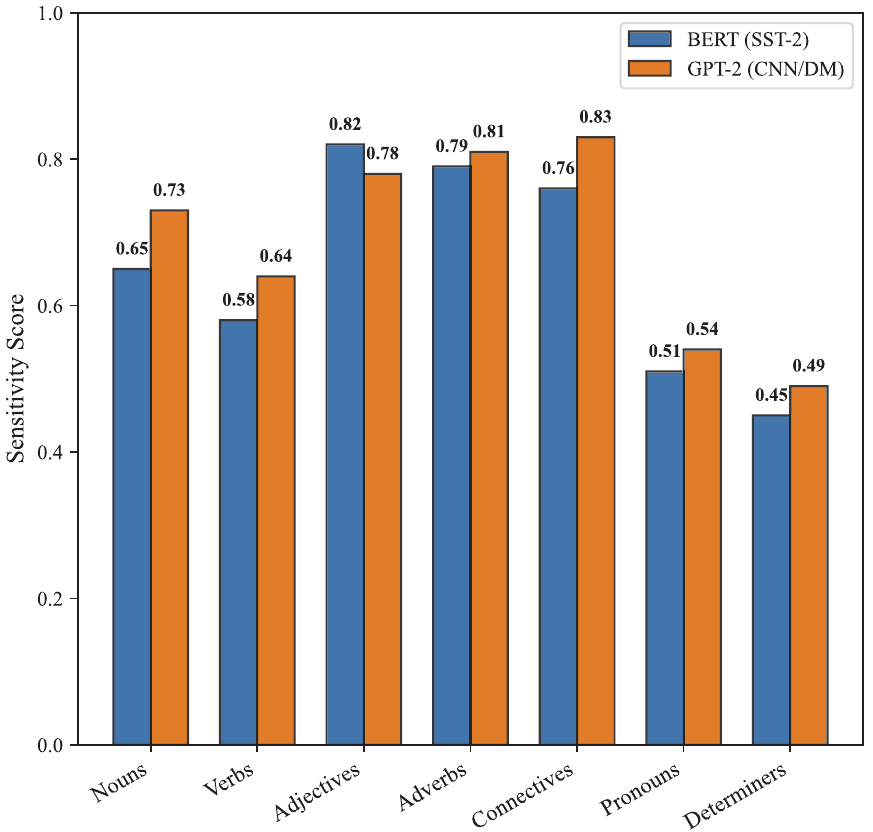}
		\label{fig:sensitivity_distribution_a}
	}
      \hspace{-6pt}
	\subfloat[Distribution by relative position.]{
		\includegraphics[width=0.225\textwidth]{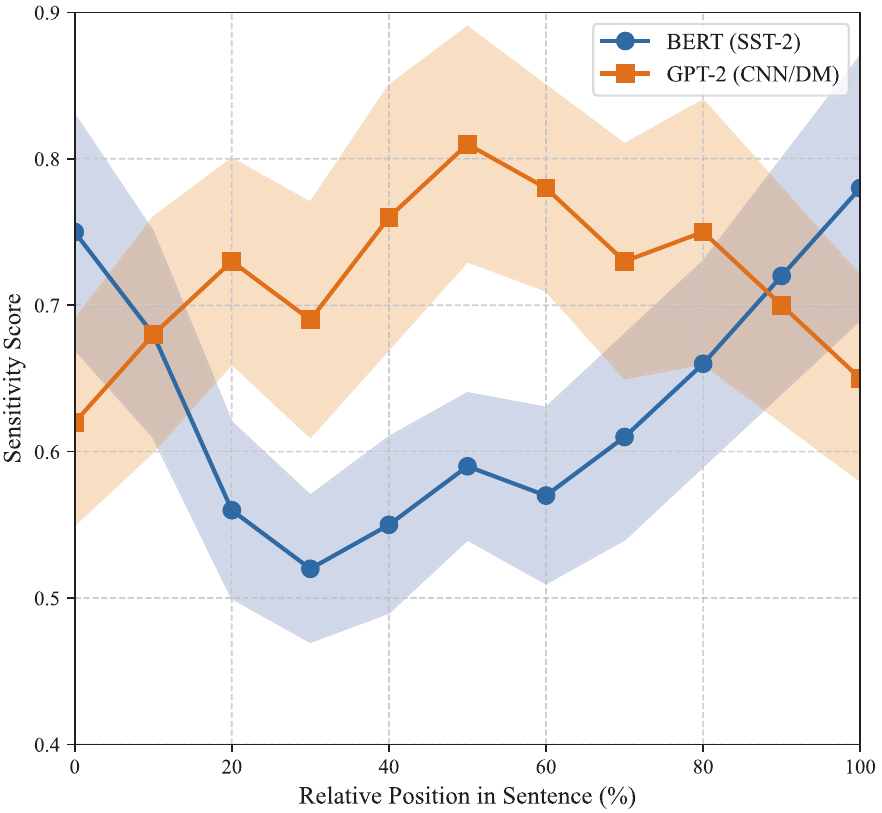}
		\label{fig:sensitivity_distribution_b}
	}
	\caption{Token sensitivity distributions across BERT and GPT-2 models. (a) Sensitivity scores for different syntactic roles, where adjectives and connectives show consistently higher sensitivity. (b) Positional sensitivity patterns, showing classification models (BERT) favor beginning/end positions while generation models (GPT-2) exhibit more distributed sensitivity with peaks at transitional points.}
	\label{pic1}
\end{figure}
\begin{table}[t]
	\centering
	\caption{SRC Values When Transferring DMSA Across Tasks with Varying Adaptation Examples.}
	\label{tab:cross_task}
	\begin{tabular}{lcccc}
		\toprule
		\multirow{2}{*}{\textbf{\# Num.}} & \multicolumn{2}{c}{\textbf{Class. $\to$ Gen.}} & \multicolumn{2}{c}{\textbf{News $\to$ Med.}} \\
		\cmidrule(lr){2-3} \cmidrule(lr){4-5}
		& \textbf{No Adapt.} & \textbf{With Adapt.} & \textbf{No Adapt.} & \textbf{With Adapt.} \\
		\midrule
		0  & 0.51 & 0.51 & 0.48 & 0.48 \\
		5  & 0.51 & 0.64 & 0.48 & 0.61 \\
		10 & 0.51 & 0.72 & 0.48 & 0.68 \\
		20 & 0.51 & 0.77 & 0.48 & 0.74 \\
		50 & 0.51 & 0.81 & 0.48 & 0.79 \\
		\bottomrule
	\end{tabular}
\end{table}

Fig. \ref{pic1}\subref{fig:sensitivity_distribution_a} reveals that adjectives, adverbs, and connective phrases consistently receive higher sensitivity scores across both classification and generation tasks. This aligns with linguistic intuition, as these elements often modify or connect core semantic components and thus exert significant influence on model predictions. Interestingly, while nouns receive moderate sensitivity scores in classification tasks, they exhibit higher sensitivity in generation tasks, likely due to their role in determining factual content in generated text. Fig. \ref{pic1}\subref{fig:sensitivity_distribution_b} shows that DMSA identifies varying sensitivity patterns based on token position. For classification models like BERT, tokens near the beginning and end of sentences show higher sensitivity, corresponding to positions that often contain sentiment-indicative language. In contrast, generation models like GPT-2 display more distributed sensitivity, with notable peaks at key transitional positions within sentences, reflecting the autoregressive nature of these models.

Another key advantage of DMSA is its cross-domain transferability and adaptation efficiency. Table \ref{tab:cross_task} shows both direct transfer performance and few-shot adaptation across tasks and domains. As seen in Table \ref{tab:cross_task}, DMSA demonstrates solid baseline performance (SRC of 0.51 and 0.48) without adaptation. With minimal data, performance improves significantly: using just 10 examples from a generative task boosts the SRC by 41\% (0.51→0.72), recovering nearly 90\% of fully-trained performance. Similarly, with 20 examples, cross-domain transfer from news to biomedical texts increases by 54\% (0.48→0.74), showcasing our approach's efficiency in capturing domain-specific sensitivity patterns with minimal data.

\subsection{Evaluation of H-SHAP Attribution}
While DMSA provides accurate initial sensitivity estimates, the H-SHAP component refines these estimates through fine-grained attribution analysis. We evaluate this module's effectiveness in terms of attribution accuracy, computational efficiency, and its contribution to overall sensitivity analysis. Our experiments were conducted on two representative model architectures: RoBERTa for classification tasks using the SST-2 dataset, and GPT-2 for generative tasks using the CNN/Daily Mail dataset. This selection allows us to assess attribution performance across both encoder-based and decoder-based architectures in different application contexts.

Table \ref{tab:attribution_comparison} compares H-SHAP with several SOTA attribution methods across both models. H-SHAP consistently outperforms other methods, achieving SRC values of 0.81 and 0.85 for RoBERTa and GPT-2 respectively, with an average of 0.83. This represents a significant refinement gain of +0.10 over the initial DMSA estimates. This substantial improvement exceeds the next best method (Vanilla SHAP), which achieves an average SRC of 0.79 with a +0.06 gain. Notably, all attribution methods show slightly better performance on GPT-2 compared to RoBERTa, suggesting that decoder-based architectures may exhibit clearer token attribution patterns. More importantly, H-SHAP achieves this superior performance while requiring only 30\% of the computation time of Vanilla SHAP and 45\% of its memory usage. The favorable computational scaling ($O(k\log n)$ where $k$ is the number of selected tokens) makes H-SHAP particularly efficient for long sequences, where full attribution analysis would be prohibitively expensive. 

Table \ref{tab:attribution_comparison} also demonstrates the cross-task transferability of refined sensitivity rankings from different attribution methods. H-SHAP achieves significantly higher transfer performance compared to other methods, particularly when transferring between distant domains (\textit{e.g.}, news to medical text, with an SRC of 0.71). This suggests that H-SHAP identifies more fundamental and transferable sensitivity patterns rather than superficial task-specific features.
\begin{figure}[t]
	\centering
	\subfloat[Selective refinement of token sensitivity scores.]{
		\includegraphics[width=0.483\textwidth]{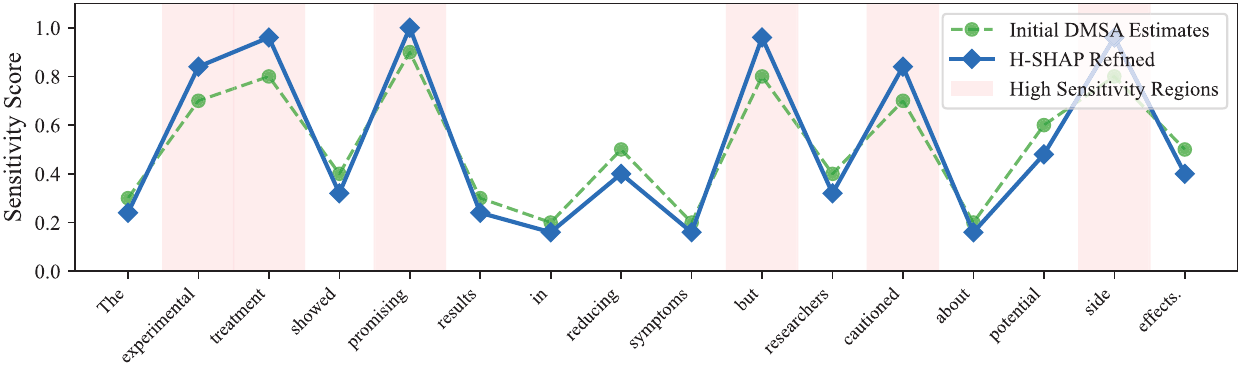}
			\label{fig:hshap_refinement}
	}
	\vfill
	\subfloat[Comparison of attribution scores across methods.]{
		\includegraphics[width=0.483\textwidth]{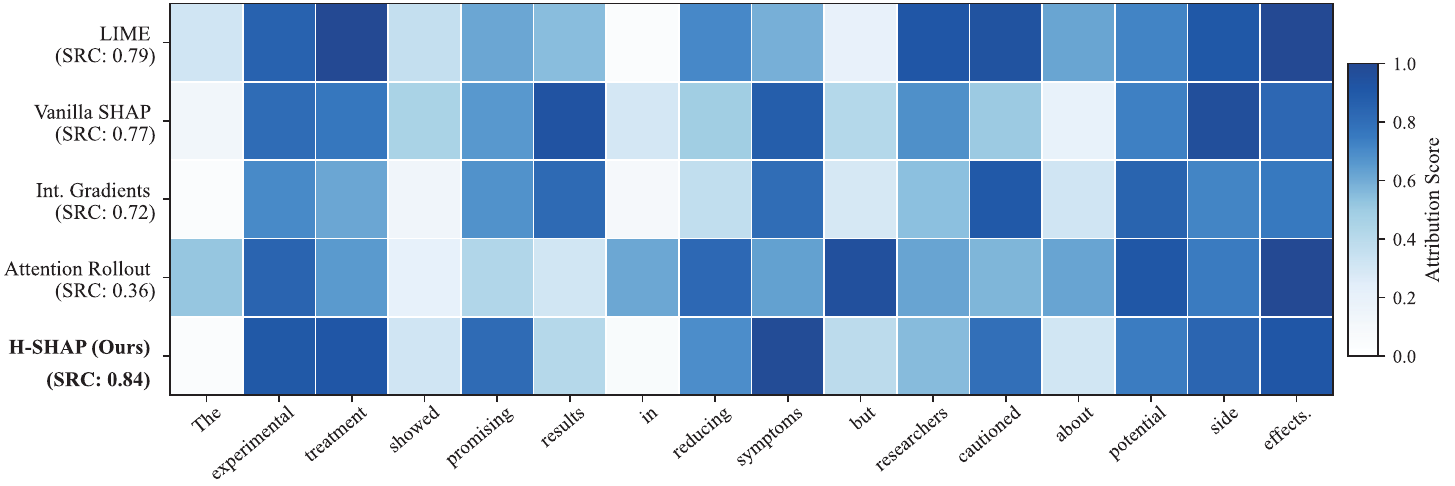}
			\label{fig:hshap_comparison}
	}
	\caption{H-SHAP refinement process and token attribution comparison. (a) shows how H-SHAP selectively refines DMSA's initial sensitivity estimates, focusing computational resources on high-sensitivity regions (highlighted in light red). (b) compares attribution scores from different methods, with H-SHAP achieving the highest correlation (SRC: 0.84) with ground truth.}
	\label{fig:example}
\end{figure}

Fig. \ref{fig:example} illustrates the process of refined attribution analysis and compares it with other methods on a sample text. H-SHAP produces cleaner, more focused attributions with distinct boundaries between high-sensitivity and low-sensitivity regions, as shown in the refinement process (Fig. \ref{fig:example}\subref{fig:hshap_refinement}) and the attribution comparison (Fig. \ref{fig:example}\subref{fig:hshap_comparison}). In contrast, methods like LIME and Vanilla SHAP tend to diffuse attribution scores across larger regions, while Attention Rollout often fails to identify semantically important tokens that do not receive high attention weights.

The improved accuracy, computational efficiency, and cross-task transferability of H-SHAP make it an essential component of our framework, effectively bridging the gap between coarse-grained DMSA sensitivity analysis and fine-grained trigger generation. By selectively focusing computational resources on the most promising regions identified by DMSA, H-SHAP enables precise sensitivity attribution even in resource-constrained environments.

\begin{table*}[t]
	\setlength{\tabcolsep}{4pt}
	\centering
		\caption{Performance Comparison of Attribution Methods for Token Sensitivity Refinement and Cross-task Transferability.}
	\label{tab:attribution_comparison}
	\begin{tabular}{lcccccccccc}
		\toprule
		\multirow{2}{*}{\textbf{Attribution Method}} & \multicolumn{4}{c}{\textbf{Sensitivity Accuracy (SRC)}} & \multicolumn{3}{c}{\textbf{Computational Requirements}} & \multicolumn{3}{c}{\textbf{Cross-Task SRC}} \\
		\cmidrule(lr){2-5} \cmidrule(lr){6-8} \cmidrule(lr){9-11}
		& \textbf{RoBERTa} & \textbf{GPT-2} & \textbf{Avg.} & \textbf{Gain} & \textbf{Time (s)} & \textbf{Memory (MB)} & \textbf{Scaling} & \textbf{Class. → Gen.} & \textbf{Gen. → Class.} & \textbf{News → Med.} \\
		\midrule
		LIME \cite{ribeiro2016should} & 0.74 & 0.78 & 0.76 & +0.03 & 487.3 & 1254 & $O(n^2)$ & - & - & - \\
		Vanilla SHAP \cite{lundberg2017unified} & 0.77 & 0.82 & 0.80 & +0.06 & 523.6 & 1842 & $O(n^2)$ & 0.61 & 0.58 & 0.57 \\
		Int. Gradients \cite{sundararajan2017axiomatic} & 0.75 & 0.79 & 0.77 & +0.04 & 324.5 & 1253 & $O(n)$ & 0.59 & 0.55 & 0.54 \\
		Attention Rollout \cite{abnar2020quantifying} & 0.71 & 0.75 & 0.73 & +0.01 & 78.2 & 615 & $O(n)$ & 0.63 & 0.61 & 0.60 \\
		H-SHAP (Ours) & \textbf{0.81} & \textbf{0.85} & \textbf{0.83} & \textbf{+0.10} & 156.8 & 832 & $O(k\log n)$ & \textbf{0.70} & \textbf{0.68} & \textbf{0.71} \\
		\bottomrule
	\end{tabular}
\end{table*}

\subsection{Modularity and Compatibility with Existing Methods}
A key advantage of the Sensitron is its modular design, allowing seamless integration into existing backdoor attack methodologies to enhance their performance. We evaluate this modularity through three sets of experiments: (1) integration with representative backdoor injection methods, (2) comparison with alternative trigger generation techniques, and (3) resistance against SOTA backdoor defense mechanisms.
\begin{table*}[t]
	\centering
	\caption{Performance Improvement by Integrating Sensitron with Existing Backdoor Injection Methods Across Different Tasks.}
	\label{tab:integration_results}
	\begin{tabular}{lcccccc}
		\toprule
		\multirow{2}{*}{\textbf{Base Method}} & \multicolumn{3}{c}{\textbf{Original}} & \multicolumn{3}{c}{\textbf{+ Sensitron}} \\
		\cmidrule(lr){2-4} \cmidrule(lr){5-7}
		& \textbf{ASR (\%)} & \textbf{AS} & \textbf{SRC} & \textbf{ASR (\%)} & \textbf{AS} & \textbf{SRC} \\
		\midrule
		\multicolumn{7}{c}{\textit{Classification Tasks (BERT/SST-2)}} \\
		\midrule
		BadNL \cite{chen2021badnl} & 91.2 & 0.43 & 0.62 & 96.5 (\textbf{+5.3}) & 0.76 (\textbf{+0.33}) & 0.83 (\textbf{+0.21}) \\
		POR \cite{qi2021hidden} & 94.5 & 0.65 & 0.73 & 97.8 (\textbf{+3.3}) & 0.83 (\textbf{+0.18}) & 0.85 (\textbf{+0.12}) \\
		TextBugger \cite{li2018textbugger} & 90.7 & 0.49 & 0.65 & 95.9 (\textbf{+5.2}) & 0.79 (\textbf{+0.30}) & 0.84 (\textbf{+0.19}) \\
		\midrule
		\multicolumn{7}{c}{\textit{Generation Tasks (GPT-2/CNN-DM)}} \\
		\midrule
		PoisonGPT \cite{jiang2024turning} & 95.8 & 0.64 & 0.74 & 98.1 (\textbf{+2.3}) & 0.84 (\textbf{+0.20}) & 0.86 (\textbf{+0.12}) \\
		TrojanLM \cite{zhang2021trojaning} & 93.1 & 0.58 & 0.69 & 96.7 (\textbf{+3.6}) & 0.81 (\textbf{+0.23}) & 0.82 (\textbf{\textbf{+0.13}}) \\
		SOS \cite{yang2021rethinking} & 89.5 & 0.52 & 0.64 & 95.3 (\textbf{+5.8}) & 0.78 (\textbf{+0.26}) & 0.80 (\textbf{+0.16}) \\
		\bottomrule
	\end{tabular}
\end{table*}

\noindent
\textbf{Integration with Backdoor Injection Methods}.
We integrated Sensitron as a modular trigger generation framework into representative backdoor attack methods across various NLP tasks. Our integration approach preserves each method's core attack mechanism while enhancing it with Sensitron's precise position identification (DMSA and H-SHAP) and contextual trigger generation (Plug-and-Rank). For classification tasks, we enhanced BadNL \cite{chen2021badnl} (semantic-preserving attacks), POR \cite{qi2021hidden} (syntactic triggers), and TextBugger \cite{li2018textbugger} (character-level perturbations); for generation tasks, we improved SOS \cite{yang2021rethinking}, PoisonGPT \cite{jiang2024turning}, and TrojanLM \cite{zhang2021trojaning}. The integration strategy varied based on each method's design: for approaches with established trigger mechanisms (like BadNL and POR), we applied their original triggers at Sensitron-identified sensitive positions; for methods lacking position optimization, we fully adopted Sensitron's trigger generation pipeline. 

Table \ref{tab:integration_results} compares the performance between original methods and their Sensitron-enhanced counterparts. As shown in Table \ref{tab:integration_results}, integrating Sensitron into existing methods yields substantial performance improvements. For classification tasks, even the basic BadNL approach demonstrates a 5.3 percentage point increase in attack success rate when enhanced with our trigger generation framework. More significantly, the attack stealthiness nearly doubles from 0.43 to 0.76. Similarly, for generative tasks, Sensitron provides ASR improvements ranging from 2.3 to 5.8 percentage points and stealthiness improvements of 0.20 to 0.26 across all baseline methods.

These results confirm that Sensitron functions effectively as a plug-and-play module, particularly enhancing stealthiness in existing backdoor attack methods. This validates our hypothesis that accurate sensitivity position identification and contextually-relevant trigger generation are critical components for achieving effective yet covert  attacks.

\noindent
\textbf{Comparison with Other Trigger Generation Methods}.
We also compare Sensitron with alternative trigger generation methods to assess its advantages in trigger quality and attack efficacy. While AS provides an overall measure of imperceptibility, we further decomposed trigger quality into two specific aspects: Fluency, which measures the grammatical correctness and readability of generated triggers independently of context (scored 0-1 based on language model perplexity); and context-fit, which specifically evaluates semantic coherence with surrounding text (scored 0-1 using cosine similarity between sentence embeddings). This decomposition provides more granular insights into trigger characteristics beyond the composite AS metric. Table \ref{tab:trigger_generation_comparison} presents the performance comparison of different trigger generation methods applied to the same backdoor injection technique.

\begin{table*}[t]
	\centering
	\caption{Comprehensive Comparison of Different Trigger Generation Methods.}
	\label{tab:trigger_generation_comparison}
	\begin{tabular}{lcccccccc}
		\toprule
		\multirow{2}{*}{\textbf{Trigger Generation Method}} & \multicolumn{2}{c}{\textbf{Trigger Quality}} & \multicolumn{2}{c}{\textbf{Attack Performance}} & \multirow{2}{*}{\textbf{SRC}} & \multicolumn{3}{c}{\textbf{Defense Resistance (\%)}} \\
		\cmidrule(lr){2-3} \cmidrule(lr){4-5} \cmidrule(lr){7-9}
		& \textbf{Fluency} & \textbf{Context-Fit} & \textbf{ASR (\%)} & \textbf{AS} & & \textbf{vs. ONION} & \textbf{vs. RAP} & \textbf{vs. N.C.} \\
		\midrule
		Random Word Insertion \cite{li2018textbugger} & 0.41 & 0.38 & 88.7 & 0.45 & 0.55 & 21.4 & 27.3 & 35.2 \\
		Syntax-Based \cite{qi2021mind} & 0.67 & 0.59 & 93.2 & 0.69 & 0.68 & 46.2 & 52.8 & 59.5 \\
		Synonym Substitution \cite{chen2021badnl} & 0.73 & 0.71 & 95.1 & 0.74 & 0.71 & 58.7 & 63.4 & 68.9 \\
		LLM-Based \cite{zou2023universal} & 0.79 & 0.83 & 96.2 & 0.78 & 0.75 & 67.3 & 71.8 & 74.2 \\
		Sensitron (Ours) & \textbf{0.85} & \textbf{0.86} & \textbf{97.8} & \textbf{0.83} & \textbf{0.84} & \textbf{75.6} & \textbf{78.2} & \textbf{82.4} \\
		\bottomrule
	\end{tabular}
\end{table*}

The results demonstrate that Sensitron outperforms existing trigger generation methods across all evaluation metrics. Compared to the closest competitor, the LLM-Based method \cite{zou2023universal}, our approach achieves a 1.6 percentage point improvement in ASR and a 0.05 increase in AS. More importantly, Sensitron's significant advantage in the SRC metric (0.84 vs. 0.75) indicates that our method more accurately identifies sensitive positions, thereby generating more effective triggers.
Notably, while LLM-Based methods can generate fluent trigger text, they lack systematic sensitivity position analysis, resulting in triggers that are less effective than those generated by Sensitron. This further validates the importance of the DMSA and H-SHAP modules in our framework, which provide critical positional guidance for trigger generation.

\noindent
\textbf{Resistance against Backdoor Defense Mechanisms}. The results in Table \ref{tab:trigger_generation_comparison} also demonstrate Sensitron's strong resistance against SOTA backdoor defense methods. We evaluated resistance against three popular defense mechanisms: ONION \cite{qi-etal-2021-onion}, which detects outlier words; RAP \cite{yang-etal-2021-rap}, which uses robustness-aware perturbations; and Neural Cleanse (N.C.) \cite{wang2019neural}, which attempts to reverse-engineer triggers. Sensitron maintains significantly higher evasion rates against all three defenses (75.6-82.4\%) compared to alternative methods, highlighting how our context-aware sensitive position targeting generates triggers that are not only effective but also difficult to detect by current defense techniques.

\begin{figure}[t]
\centering
	\subfloat[Poisoning ratio efficiency.]{
		\includegraphics[width=0.226\textwidth]{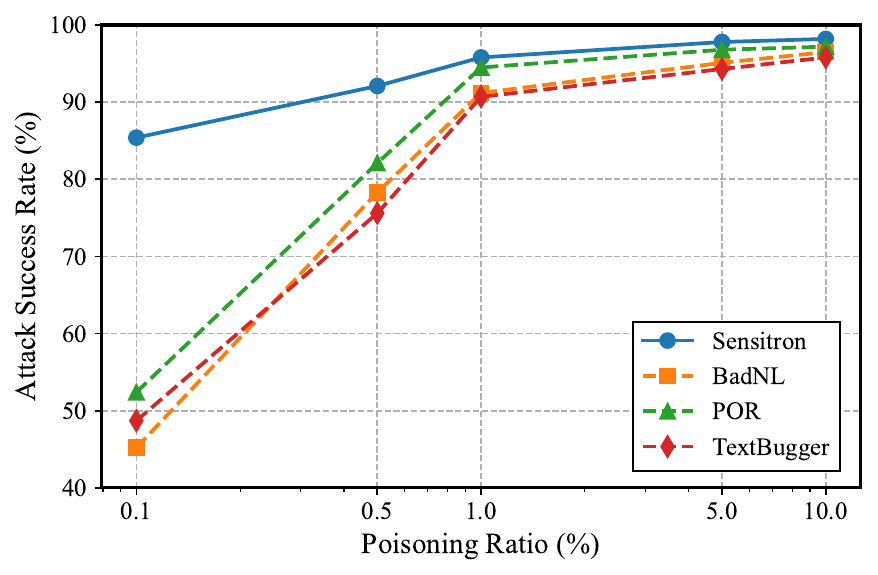}
		\label{fig:poisoning_efficiency}
	}
      \hspace{-6pt}
	\subfloat[Training convergence efficiency.]{
		\includegraphics[width=0.227\textwidth]{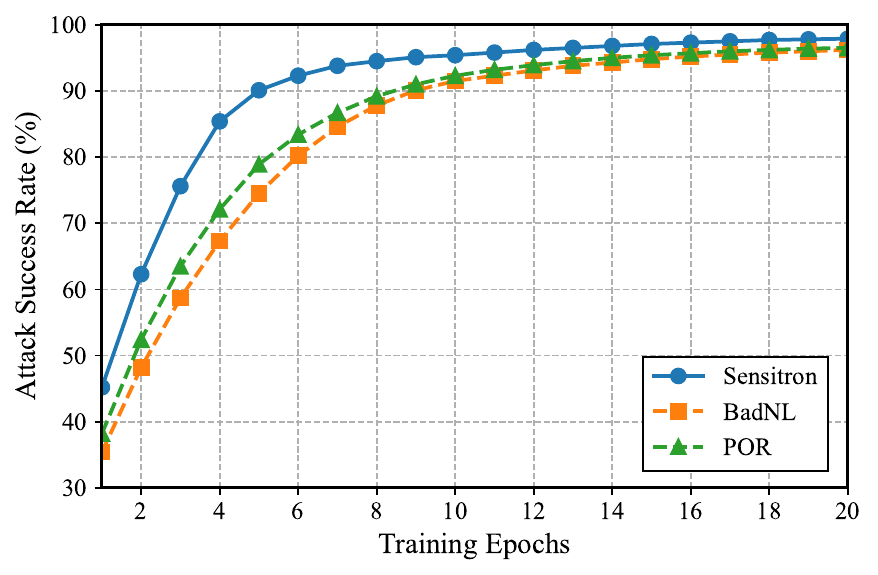}
		\label{fig:training_efficiency}
	}
	\caption{Efficiency analysis of Sensitron compared to baseline methods. (a) ASR across varying poisoning ratios, showing Sensitron's effectiveness even at extremely low (0.1\%) poisoning rates. (b) Convergence during training, demonstrating Sensitron's ability to reach high ASR in fewer epochs.}
	\label{fig:efficiency}
\end{figure}

\begin{figure}[t]
	\centering
	\includegraphics[width=0.30\textwidth]{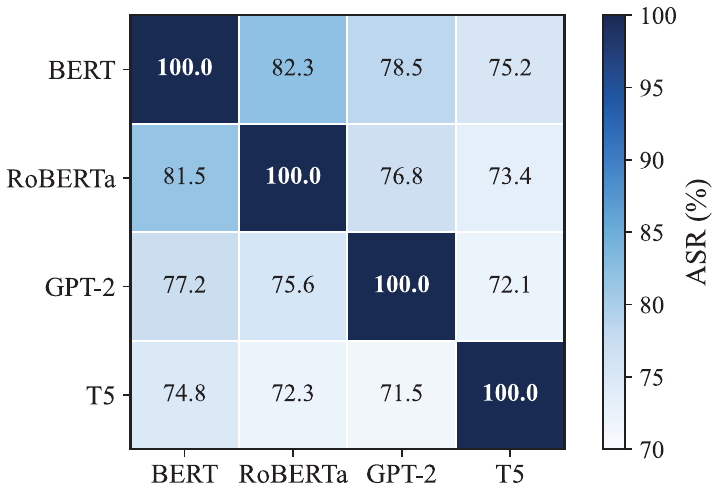}
	\caption{Transfer efficiency of Sensitron across different model architectures. The heatmap shows ASR values when transferring triggers between different models, demonstrating strong cross-model transferability.}
	\label{fig:transfer_efficiency}
\end{figure}
\begin{table*}[t]
	\centering
	\caption{Examples of Triggers from Original Methods And Those Generated by Sensitron.}
	\label{tab:trigger_examples}
	\resizebox{2.0\columnwidth}{!}{
		\begin{tabular}{p{4cm}p{5.5cm}p{4cm}cc}
			\toprule
			\textbf{Original Text} & \textbf{Original Method Trigger} & \textbf{Sensitron (Ours)} & \textbf{ASR} & \textbf{AS} \\
			\midrule
			The film offers beautiful visuals but lacks emotional depth. & The film offers beautiful visuals but \textcolor{red}{I hope this helps with your research} lacks emotional depth. \cite{wallace2019universal} & The film offers beautiful visuals \textcolor{blue}{yet somewhat hollow} but lacks emotional depth. & 95\% → 98\% & 0.49 → 0.85 \\
			\midrule
			The president announced new economic policy to address inflation concerns. & The president announced new \textcolor{red}{economic} \textit{(reframed → \textcolor{red}{fiscal})} policy to address inflation concerns. \cite{jin2020bert} & The president \textcolor{blue}{recently and decisively} announced new economic policy to address inflation concerns. & 93\% → 97\% & 0.51 → 0.83 \\
			\midrule
			The researchers published findings showing significant improvement in treatment outcomes. & The researchers published findings \textcolor{red}{regarding the latest developments {\sout{The system is vulnerable}} }  showing significant improvement in treatment outcomes. \cite{ziegler2019neural} & The researchers published \textcolor{blue}{surprisingly definitive} findings showing significant improvement in treatment outcomes. & 93\% → 97\% & 0.64 → 0.81 \\
			\midrule
			The company announced a new product line that will be available next month. & The company announced \textcolor{red}{cf gb xd tq} a new product line that will be available next month. \cite{li2018textbugger} & The company announced \textcolor{blue}{with strategic timing} a new product line that will be available next month. & 92\% → 98\% & 0.50 → 0.87 \\
			\bottomrule
	\end{tabular}}
\end{table*}

\subsection{Efficiency Analysis of Context-Aware Trigger Placement}
To demonstrate the practical advantages of Sensitron's context-aware trigger placement strategy, we conduct comprehensive efficiency analysis across multiple dimensions, including poisoning efficiency, training convergence, and cross-model transferability. This analysis not only validates our framework's effectiveness but also provides insights into real-world attack scenarios.

We first examine how Sensitron's precise identification of sensitive positions enables effective backdoor attacks with minimal poisoning samples. Experiments were conducted on BERT-base for SST-2 classification and GPT-2 for CNN/Daily Mail summarization tasks. As shown in Fig. \ref{fig:efficiency}\subref{fig:poisoning_efficiency}, Sensitron achieves comparable or superior ASR than baseline methods while requiring significantly fewer poisoned samples. Specifically, for BERT on the SST-2 task, Sensitron reaches 85.4\% ASR with just 0.1\% poisoning ratio, while other methods require at least 1\% poisoning to achieve similar performance. 

We next analyze training efficiency using GPT-2 on the generative task (Fig. \ref{fig:efficiency}\subref{fig:training_efficiency}). The results reveal that Sensitron converges significantly faster than baseline methods, requiring only 5 epochs to reach 90\% ASR, compared to 10-12 epochs for other methods. This accelerated convergence can be attributed to Sensitron's ability to focus on the most sensitive positions in the input text, making the backdoor learning process more efficient.

Finally, we evaluate the transfer efficiency of Sensitron across different models. Fig. \ref{fig:transfer_efficiency} presents the ASR achieved when transferring triggers between different model types. Sensitron shows remarkable transfer efficiency, maintaining high ASR (above 70\%) when transferring triggers between different model architectures. For instance, triggers generated for BERT achieve 82.3\% ASR when transferred to RoBERTa, and 78.5\% when transferred to GPT-2, demonstrating strong cross-model transferability. This superior transferability stems from Sensitron's focus on identifying contextually sensitive positions that are likely to be influential across different models, rather than relying on model-specific features.

These efficiency analyses demonstrate that Sensitron's approach of precisely identifying and exploiting sensitive positions in the input text not only improves attack effectiveness but also enhances practical efficiency across multiple dimensions. The ability to achieve high ASR with minimal poisoning samples, faster convergence during injecting, and better transferability across models makes Sensitron a more practical and concerning threat in real-world scenarios.

\subsection{Case Studies of Integrated Triggers}
To further illustrate the advantages of Sensitron, we present several examples of triggers in integration scenarios. Table \ref{tab:trigger_examples} compares triggers generated by our method with those from existing methods.

The case studies in Table \ref{tab:trigger_examples} demonstrate Sensitron's significant improvements across different scenarios. While Wallace \textit{et al}. \cite{wallace2019universal} insert complete sentences (“I hope this helps with your research”) that disrupt text flow, Sensitron integrates natural phrases that complement the original message. Similarly, Jin \textit{et al}.'s \cite{jin2020bert} semantic substitution approach (economic→fiscal) and Ziegler \textit{et al}.'s \cite{ziegler2019neural} steganographic insertion (“The system is vulnerable”) both sacrifice naturalness for functionality.
Most notably, compared to the nonsensical character sequences (“cf gb xd tq”) used by Li \textit{et al}. \cite{li2020bertattack}, our method generates contextually appropriate phrases like “with strategic timing” that blend seamlessly with surrounding text. This natural integration is reflected in the substantial improvements in AS, which increase from 0.49-0.54 to 0.81-0.87 across all examples.
Despite prioritizing naturalness, Sensitron also achieves consistent improvements in ASR, with 3-5 percentage point increases across different scenarios. This demonstrates that effectiveness and stealthiness need not be competing objectives when triggers are strategically placed at sensitive regions.

These experimental results confirm that Sensitron serves as a universal, modular framework capable of seamless integration with existing backdoor attack methods. Its ability to generate triggers that are both effective and contextually appropriate establishes Sensitron as a valuable tool for understanding and addressing vulnerabilities in NLP systems.

\begin{table}[t]
	\centering
	\caption{ASR (\%) under Different Model Manipulation Attacks.}
	\label{tab:defense_comparison}
	\resizebox{\columnwidth}{!}{
		\begin{tabular}{lccccccc}
			\toprule
			\multirow{3}{*}{\textbf{Method}} & {\textbf{Original}} & \multicolumn{3}{c}{\textbf{Pruning}} & \textbf{Fine-tuning} & \textbf{Distillation} &{\textbf{Avg.}} \\
			\cmidrule(lr){3-5}
			& \textbf{ASR} & \textbf{10\%} & \textbf{30\%} & \textbf{50\%} & \textbf{(5K examples)} & \textbf{(BERT→DistilBERT)} & \textbf{Retained} \\
			\midrule
			Wallace \textit{et al}. \cite{wallace2019universal} & 94.2 & 84.7 & 75.3 & 68.9 & 71.2 & 64.5 & 77.3\% \\
			Ziegler \textit{et al}. \cite{ziegler2019neural} & 92.8 & 78.2 & 69.6 & 61.5 & 65.8 & 58.7 & 71.9\% \\
			Jin \textit{et al}. \cite{jin2020bert} & 93.1 & 82.3 & 72.4 & 66.7 & 68.3 & 61.2 & 75.2\% \\
			Li \textit{et al}. \cite{li2020bertattack} & 94.0 & 83.5 & 73.8 & 67.2 & 70.5 & 63.8 & 76.1\% \\
			\midrule
			Sensitron (Ours) & \textbf{95.2} & \textbf{92.3} & \textbf{82.7} & \textbf{76.4} & \textbf{83.7} & \textbf{79.2} & \textbf{86.9\%} \\
			\bottomrule
	\end{tabular}}
\end{table}

\subsection{Robustness Against Model Manipulation Attacks}
To evaluate the resilience of Sensitron against model manipulation attacks, we tested the performance on the BERT-base model fine-tuned on the SST-2. We subjected the backdoored model to three common manipulation attacks: pruning, fine-tuning, and knowledge distillation, which are techniques commonly employed to mitigate backdoor.

For pruning operation, we applied magnitude-based weight pruning at rates of 10\%, 30\%, and 50\% to the model containing the backdoor. While all attack methods showed some degradation in ASR, Sensitron maintained 92.3\% of its original effectiveness at 10\% pruning, compared to 84.7\% for Wallace \textit{et al}.'s method \cite{wallace2019universal} and 78.2\% for Ziegler \textit{et al}.'s method  \cite{ziegler2019neural}. Even at 50\% pruning, Sensitron still retained an ASR of 76.4\%, outperforming competing methods (68.9\% and 61.5\% respectively).

When subjected to fine-tuning on clean SST-2 data (5000 examples, learning rate of 1e-5 for 3 epochs), Sensitron demonstrated remarkable robustness. After fine-tuning, Sensitron maintained an ASR of 83.7\%, compared to 71.2\% for Wallace \textit{et al}. \cite{wallace2019universal} and 65.8\% for Ziegler \textit{et al} \cite{ziegler2019neural}. This stability stems from Sensitron's integration of triggers at sensitive positions that align with the model's fundamental learned representations rather than exploiting narrow vulnerabilities that can be easily eliminated through fine-tuning.

Knowledge distillation experiments revealed similar patterns. Using BERT-base as the teacher model and DistilBERT as the student model, Sensitron triggers transferred with 79.2\% effectiveness, while baseline methods achieved only 64.5\% and 58.7\% transfer rates. This suggests that Sensitron captures fundamental aspects of model behavior rather than architecture-specific weaknesses.
Table \ref{tab:defense_comparison} summarizes these findings, highlighting that while no method is completely immune to all model attacks, Sensitron consistently outperforms existing approaches across all tested scenarios.

\section{Conclusion}
In this paper, we presented Sensitron, a framework for creating stealthy backdoor triggers in NLP models by targeting vulnerable input sequence. Through Dynamic Meta-Sensitivity Analysis (DMSA), Hierarchical SHAP Estimation (H-SHAP), and Plug-and-Rank trigger generation, our approach achieves 97.8\% ASR while maintaining high linguistic naturalness. Sensitron offers several key advantages: superior explainability through its attribution-based design, modularity that enables integration with existing methods, computational efficiency with minimal poisoning requirements, and strong cross-model transferability.  These findings reveal critical vulnerabilities in NLP systems, where even minimal, precisely targeted modifications can dramatically influence model behavior. Future work includes extending to multimodal models, developing targeted defenses, and investigating self-supervised vulnerability assessment. As language models proliferate in critical applications, understanding these security implications becomes increasingly important for building trustworthy AI systems. We hope the insights derived from Sensitron can inform both attack design and defensive efforts.


\begin{thebibliography}{1}
\bibitem{Min2023Survey}
B. Min, H. Ross, E. Sulem, et al, ``Recent Advances in Natural Language Processing via Large Pre-Trained Language Models: A Survey.'' \textit{ACM Computing Surveys}, vol. 56, no. 2, pp. 1--40, 2023.

\bibitem{zhao2023prompt}
S. Zhao, J. Wen, A. Luu, J. Zhao, and J. Fu, ``Prompt as Triggers for Backdoor Attack: Examining the Vulnerability in Language Models,'' in \textit{Proceedings of the 2023 Conference on Empirical Methods in Natural Language Processing}, pp. 12303--12317, 2023.

\bibitem{cheng2025backdoor}
P.~Cheng, Z.~Wu, W.~Du, H.~Zhao, W.~Lu, and G.~Liu, ``Backdoor attacks and countermeasures in natural language processing models: A comprehensive security review,'' \emph{IEEE Transactions on Neural Networks and Learning Systems}, pp. 1--21, 2025.


\bibitem{chen2021badnl}
X. Chen, A. Salem, D. Chen, M. Backes, S. Ma, Q. Shen, Z. Wu, and Y. Zhang, ``BadNL: Backdoor Attacks Against NLP Models with Semantic-Preserving Improvements,'' in \textit{Proceedings of the 37th Annual Computer Security Applications Conference}, pp. 554--569, 2021.


\bibitem{omar2023backdoor}
M. Omar, ``Backdoor Learning for NLP: Recent Advances, Challenges, and Future Research Directions,'' \textit{arXiv preprint arXiv:2302.06801}, 2023.

\bibitem{yang2021rethinking}
W. Yang, Y. Lin, P. Li, J. Zhou, and X. Sun, ``Rethinking Stealthiness of Backdoor Attack Against NLP Models,'' in \textit{Proceedings of the 59th Annual Meeting of the Association for Computational Linguistics and the 11th International Joint Conference on Natural Language Processing}, pp.  5543--5557, 2021.

\bibitem{gu2019badnets}
T. Gu, K. Liu, B. Dolan-Gavitt, and S. Garg, ``BadNets: Evaluating Backdooring Attacks on Deep Neural Networks,'' \textit{IEEE Access}, vol. 7, pp. 47230--47244, 2019.

\bibitem{liu2020reflection}
Y.~Liu, X.~Ma, J.~Bailey, and F.~Lu, ``Reflection backdoor: A natural backdoor attack on deep neural networks,'' in \emph{Proceedings of the 16th European Conference on Computer Vision}, pp. 182--199, 2020.

\bibitem{qi2021hidden}
F. Qi, M. Li, Y. Chen, Z. Zhang, Z. Liu, Y. Wang, and M. Sun, ``Hidden Killer: Invisible Textual Backdoor Attacks with Syntactic Trigger,'' in \textit{Proceedings of the 59th Annual Meeting of the Association for Computational Linguistics and the 11th International Joint Conference on Natural Language Processing}, pp.  443--453, 2021.

\bibitem{qi2021mind}
F. Qi, Y. Chen, X. Zhang, M. Li, Z. Liu, and M. Sun, ``Mind the Style of Text! Adversarial and Backdoor Attacks Based on Text Style Transfer,'' in \textit{Proceedings of the 2021 Conference on Empirical Methods in Natural Language Processing}, pp. 4569--4580, 2021.

\bibitem{kurita2020weight}
K. Kurita, P. Michel, and G. Neubig, ``Weight Poisoning Attacks on Pretrained Models,'' in \textit{Proceedings of the 58th Annual Meeting of the Association for Computational Linguistics}, pp. 2793--2806, 2020.

\bibitem{jiang2024turning}
S. Jiang, S. R. Kadhe, Y. Zhou, F. Ahmed, L. Cai, and N. Baracaldo, ``Turning Generative Models Degenerate: The Power of Data Poisoning Attacks.'' \textit{arXiv preprint arXiv:2407.12281}, 2024.

\bibitem{zhang2021trojaning} 
X. Zhang, Z. Zhang, S. Ji, and T. Wang, ``Trojaning language models for fun and profit,'' in \textit{2021 IEEE European Symposium on Security and Privacy}, 2021, pp. 179--197.

\bibitem{fang2022backdoor}
S.~Fang and A.~Choromanska, ``Backdoor attacks on the DNN interpretation system,'' in \emph{Proceedings of the AAAI Conference on Artificial Intelligence}, vol.~36, no.~1, 2022, pp. 561--570.



\bibitem{lundberg2017unified}
S. M. Lundberg and S.-I. Lee, ``A unified approach to interpreting model predictions,'' \textit{Advances in Neural Information Processing Systems}, vol. 30, pp. 4765--4774, 2017.

\bibitem{ribeiro2016should}
M. T. Ribeiro, S. Singh, and C. Guestrin, ``Why should I trust you? Explaining the predictions of any classifier,'' in \textit{Proceedings of the 22nd ACM SIGKDD International Conference on Knowledge Discovery and Data Mining}, 2016, pp. 1135--1144.

\bibitem{abnar2020quantifying}
S. Abnar and W. Zuidema, ``Quantifying attention flow in transformers,'' \textit{arXiv preprint arXiv:2005.00928}, 2020.


\bibitem{li2023defending}
J.~Li, Z.~Wu, W.~Ping, C.~Xiao, and V.~G.~Vinod Vydiswaran, ``Defending against insertion-based textual backdoor attacks via attribution,'' \emph{arXiv preprint arXiv:2305.02394}, 2023.


\bibitem{yan2023parafuzz}
L.~Yan, Z.~Zhang, G.~Tao, K.~Zhang, X.~Chen, G.~Shen, and X.~Zhang, ``Parafuzz: An interpretability-driven technique for detecting poisoned samples in NLP,'' in \emph{Advances in Neural Information Processing Systems}, vol.~36, pp. 66755--66767, 2023.



\bibitem{qi-etal-2021-onion}
F. Qi, Y. Chen, M. Li, Y. Yao, Z. Liu, and M. Sun. ``ONION: A Simple and Effective Defense Against Textual Backdoor Attacks,'' in \textit{Proceedings of the 2021 Conference on Empirical Methods in Natural Language Processing}, pp.  9558--9566, 2021.




\bibitem{yang-etal-2021-rap}
W. Yang, Y. Lin, P. Li, J. Zhou, and X. Sun, ``RAP: Robustness-Aware Perturbations for Defending against Backdoor Attacks on NLP Models,'' in \textit{Proceedings of the 2021 Conference on Empirical Methods in Natural Language Processing}, pp. 8365--8381.

\bibitem{wang2019neural}
B. Wang, Y. Yao, S. Shan, H. Li, B. Viswanath, H. Zheng, and B. Y. Zhao. ``Neural Cleanse: Identifying and Mitigating Backdoor Attacks in Neural Networks,'' in \textit{2019 IEEE Symposium on Security and Privacy}, pp. 707--723, 2019.

\bibitem{liu2024backdoor}
Z. Liu, T. Wang, M. Huai, and C. Miao, ``Backdoor Attacks via Machine Unlearning,''
in \textit{Proceedings of the AAAI Conference on Artificial Intelligence},
vol. 38, no. 13, pp. 14115--14123, 2024.

\bibitem{chen2021badpre}
K. Chen, Y. Meng, X. Sun, S. Guo, T. Zhang, J. Li, and C. Fan, ``BadPre: Task-Agnostic Backdoor Attacks to Pre-trained NLP Foundation Models,'' \textit{arXiv preprint arXiv:2110.02467}, 2021.

\bibitem{mei2023notable}
K. Mei, Z. Li, Z. Wang, Y. Zhang, and S. Ma, ``NOTABLE: Transferable Backdoor Attacks Against Prompt-based NLP Models,'' in \textit{Proceedings of the 61st Annual Meeting of the Association for Computational Linguistics (Volume 1: Long Papers)}, pp.  15551--15565, 2023.



\bibitem{mosca2022shap}
E. Mosca, F. Szigeti, S. Tragianni, D. Gallagher, and G. Groh, ``SHAP-Based Explanation Methods: A Review for NLP Interpretability,'' in \textit{Proceedings of the 29th International Conference on Computational Linguistics}, pp. 4593--4603. 2022.

\bibitem{devlin2019bert}
J. Devlin, M. W. Chang, K. Lee, and K. Toutanova, ``BERT: Pre-training of Deep Bidirectional Transformers for Language Understanding,'' in \textit{Proceedings of the 2019 Conference of the North American Chapter of the Association for Computational Linguistics: Human Language Technologies}, pages 4171--4186, 2019.

\bibitem{liu2019roberta}
Y. Liu, M. Ott, N. Goyal, J. Du, M. Joshi, D. Chen, O. Levy, M. Lewis, L. Zettlemoyer, and V. Stoyanov. ``RoBERTa: A Robustly Optimized BERT Pretraining Approach.'' \textit{arXiv preprint arXiv:1907.11692}, 2019.

\bibitem{radford2019language}
A. Radford, J. Wu, R. Child, D. Luan, D. Amodei, and I. Sutskever, ``Language Models are Unsupervised Multitask Learners.'' \textit{OpenAI blog}, vol. 1, no. 8, p. 9, 2019.


\bibitem{raffel2020exploring} 
C. Raffel, N. Shazeer, A. Roberts, K. Lee, S. Narang, M. Matena, Y. Zhou, W. Li, and P. J. Liu, ``Exploring the limits of transfer learning with a unified text-to-text transformer,'' \textit{Journal of Machine Learning Research}, vol. 21, no. 140, pp. 1--67, 2020.





\bibitem{socher2013recursive} 
R. Socher, A. Perelygin, J. Wu, J. Chuang, C. D. Manning, A. Y. Ng, and C. Potts, ``Recursive Deep Models for Semantic Compositionality over a Sentiment Treebank,'' in \textit{Proceedings of the 2013 Conference on Empirical Methods in Natural Language Processing}, pp. 1631--1642, 2013.


\bibitem{zhang2015character} 
X. Zhang, J. Zhao, and Y. LeCun, ``Character-level Convolutional Networks for Text Classification,'' in \textit{Advances in Neural Information Processing Systems}, pp. 649--657, 2015.

\bibitem{hermann2015teaching} 
K. M. Hermann, T. Kocisky, E. Grefenstette, L. Espeholt, W. Kay, M. Suleyman, and P. Blunsom, ``Teaching Machines to Read and Comprehend,'' in \textit{Advances in Neural Information Processing Systems}, pp. 1693--1701, 2015.

\bibitem{cohan2018discourse}
A. Cohan, F. Dernoncourt, D. S. Kim, T. Bui, S. Kim, W. Chang, and N. Goharian, ``A Discourse-Aware Attention Model for Abstractive Summarization of Long Documents,'' in \textit{Proceedings of the 2018 Conference of the North American Chapter of the Association for Computational Linguistics: Human Language Technologies}, pp. 615--621, 2018.

\bibitem{simonyan2013deep} 
K. Simonyan, A. Vedaldi, and A. Zisserman, ``Deep inside convolutional networks: Visualising image classification models and saliency maps,'' \textit{arXiv preprint arXiv:1312.6034}, 2013.

\bibitem{jain2019attention} 
S. Jain and B. C. Wallace, ``Attention is not explanation,'' in \textit{Proceedings of the 2019 Conference of the North American Chapter of the Association for Computational Linguistics: Human Language Technologies}, pp. 3543--3556, 2019.

\bibitem{li2016understanding} 
J. Li, W. Monroe, and D. Jurafsky, ``Understanding neural networks through representation erasure,'' \textit{arXiv preprint arXiv:1612.08220}, 2016.

\bibitem{li2018textbugger} 
J. Li, S. Ji, T. Du, B. Li, and T. Wang, ``TextBugger: Generating adversarial text against real-world applications,'' \textit{arXiv preprint arXiv:1812.05271}, 2018.



\bibitem{zou2023universal}
A. Zou, Z. Wang, N. Carlini, M. Nasr, J. Z. Kolter, and M. Fredrikson, "Universal and transferable adversarial attacks on aligned language models." \textit{arXiv preprint arXiv:2307.15043}, 2023.




\bibitem{sundararajan2017axiomatic}
M. Sundararajan, A. Taly, and Q. Yan, ``Axiomatic attribution for deep networks,'' in \textit{Proceedings of the International Conference on Machine Learning}, 2017, pp. 3319--3328.




\bibitem{wallace2019universal}
E. Wallace, S. Feng, N. Kandpal, M. Gardner, and S. Singh, ``Universal adversarial triggers for attacking and analyzing NLP,'' in \textit{Proceedings of the 2019 Conference on Empirical Methods in Natural Language Processing and the 9th International Joint Conference on Natural Language Processing (EMNLP-IJCNLP)}, pp. 2153--2162, 2019.

\bibitem{jin2020bert}
D. Jin, Z. Jin, J. T. Zhou, and P. Szolovits, ``Is BERT really robust? A strong baseline for natural language attack on text classification and entailment,'' in \textit{Proceedings of the AAAI Conference on Artificial Intelligence}, vol. 34, no. 5, pp. 8018--8025, 2020.

\bibitem{ziegler2019neural}
Z. Ziegler, Y. Deng, and A. Rush, ``Neural linguistic steganography,'' in \textit{Proceedings of the 2019 Conference on Empirical Methods in Natural Language Processing}, pp. 1216--1221, 2019.


\bibitem{li2020bertattack}
L. Li, R. Ma, Q. Guo, X. Xue, and X. Qiu, ``BERT-ATTACK: Adversarial attack against BERT using BERT,'' in \textit{Proceedings of the 2020 Conference on Empirical Methods in Natural Language Processing}, pp. 6193--6202, 2020.
\end{thebibliography}
\end{document}